\documentstyle[aps,prl,multicol,rotate,psfig]{revtex}
\begin{document}
\draft

\title{Coexistence of excited states in confined Ising systems }
\author{Andrzej Drzewi\'nski}
\address{Institute of Low Temperature and Structure Research, 
Polish Academy of Sciences,\\
P.O.Box 1410, 50-950 Wroc\l aw 2, Poland.}
\date{\today}
\maketitle

\begin{abstract}
Using the density-matrix renormalization-group method 
we study the two-dimensional Ising model in strip geometry.
This renormalization scheme enables us to consider the system up
to the size $300 \times \infty$ and study the influence
of the bulk magnetic field on the system at full range of temperature.
We have found out the crossover in the behavior of the correlation length
on the line of coexistence of the excited states. A detailed study of 
scaling of this line is performed. Our numerical results 
support and specify previous conclusions by Abraham, Parry, and Upton based
on the related bubble model.
\end{abstract}


\begin{multicols}{2} \narrowtext

The understanding of classical systems in confined geometries has been a challenge
for several years \cite{binder,diehl,dosch}. Among such investigated systems 
are fluids or magnets confined
between parallel walls. Studies of finite-size effects have not been limited only to
the vicinity of the critical point, but also to the first-order phase transitions,
which are less known.

In this Raport we consider the two-dimensional Ising system on a square lattice
in strip geometry ($L$ is width of the strip) with the Hamiltonian

\begin{equation}
  {\cal H}=-J \sum _{<i,j>}\sigma_i\sigma_j
	      -H\sum _i\sigma_i,
\label{ham}
\end{equation}

where the coupling $J>0$, $H$ is the bulk magnetic field and $\sigma_i = \pm 1$. The 
first sum runs 
over all nearest-neighbour pairs of sites while the second sum runs over all sites.

Even such a simple model has an interesting crossover governed by the bulk magnetic 
field \cite{PF,APU}, which value $H_x$ depends on temperature and the size of the system. 
The borderline $H_x(T;L)$ divides two different $L$ and $H$ dependencies of 
the correlation length $\xi$. Using the bubble model \cite{bubble} 
Abraham {\em et al.} found \cite{APU} that at subcritical temperatures one has 

\begin{eqnarray}
&& 1/\xi = P(T)L|H|, \,\,\,\,\,\,\,\,\,\,\,\,\quad\quad {\rm for} \,\,\,  0<|H| \leq H_x, 
\label{P} \\
&& 1/\xi = R(T)+S(T)|H|^{2/3}, \quad {\rm for} \,\,\,  |H| \geq H_x,
\label{RS}
\end{eqnarray}

where $P(T)=2m/k_B T$, $R(T)=2\sigma_0/k_B T$. Here, $S(T)$ is an unknown positive coefficient.
Furthermore, $m$ and $\sigma_0$ are the bulk spontaneous magnetization and the interfacial tension,
respectively.

The bubble model studies concluded that $H_x(T;L)$ scales towards
the first-order line according to the form \cite{APU,PS}: 

\begin{equation}
H_x(L;T) \approx A(T) L^{\alpha}+ B(T) L^{\gamma}+ C(T) L^{\delta} + \ldots,
\label{1ordline}
\end{equation}

where $\alpha=-1$, $\gamma=-5/3$, and $\delta=-7/3$.

A similar problem of higher-order corrections, but to the Kelvin equation
(the scaling of the bulk coexistence field in the presence of the parallel surface fields)
has been studied recently \cite{CDJ}. Using the density-matrix renormalization-group method 
(DMRG) \cite{white,nishino} it was found that for a large range of surface fields
and temperatures corrections are not compatible with the behavior predicted by the existing
theory \cite{KE}. It is one of reasons why we have checked out here the predictions given by
the bubble model.

\begin{figure}[b]
\centerline{
{\psfig{file=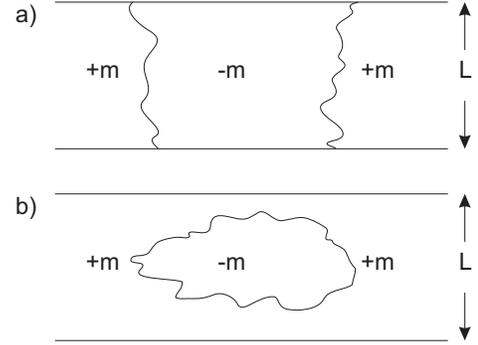,height=4.5cm}}}
\vskip 0.2truecm
\caption{The dominating configurations in strip geometry: 
a) $0<|H|<H_x$, b) $|H|>H_x$.}
\label{FIG0}
\end{figure}

Abraham {\em et al.} argued that the mentioned crossover occurs  
because the class of dominating configurations  determining the behavior 
of correlation
functions changes from a single connected loop for $|H|>H_x$ to two disconnected
closed loops $|H|<H_x$ (in cylinder geometry).
In our case, where the free boundaries are present, for $|H|<H_x$, the dominating configurations
consist of succeeding pieces of a strip with opposite magnetizations \cite{ASU}. For $|H|>H_x$ 
the most important configurations contributing to the correlation function are again
closed loops including domains of opposite magnetizations (see, Fig.1).

In order to analyze this problem beyond the bubble model, we can use the transfer-matrix (TM) 
calculations \cite{TM}. However, it is well known that to obtain satisfactory finite-size 
scaling results, one should consider large enough systems \cite{Barber}. This may, in turn, complicate
calculations or even make them impossible. To overcome this problem we have applied the DMRG method
for two-dimensional systems based on the TM approach. Providing a very efficient
algorithm for the construction of the effective transfer matrices for large $L$ this method was 
successfully employed for a number of problems (for which no exact solutions are available,
e.g. for nonvanishing bulk fields) \cite{CD,DCM,CI,Cand}. Using it we were able to analyze the system 
in full range of temperatures and the bulk magnetic field for strips of widths up to $L=300$.
For a comprehensive review of background, achievements and limitations of DMRG, 
see Ref. \cite{DMRG}.

We first calculated the free-energy levels

\begin{equation}
  f_i(H,T;L) = - \frac{k_B T}{L} \ln\Bigl(\lambda_i(H,T;L)\Bigr),
\label{fel}
\end{equation}

for $i=0,1,2,\ldots$, where $\lambda_i$ are the eigenvalues of the TM
arranged in order of decreasing magnitude. Because the inverse (longitudinal) 
correlation length can be defined as

\begin{equation}
  1/\xi(L) = \log(\lambda_0/\lambda_1),
\label{ksi}
\end{equation}

and the lowest free-energy level does not cross others,
especially important are the values of the bulk magnetic field $H_x(T;L)$, where 
the first and the second excited states cross each other. In such a case 
we can observe the crossover in the behavior of the correlation length. 

Let us first analyze the structure of the TM low-lying levels as a function of the bulk magnetic
field $H$ at fixed $T$.
At very low temperature they should behave practically in 
the same way as the ground state energy. Therefore, it is worthy first considering
the ground state properties of the system. 

Let us define the configuration of a row for the strip in the following way
$|\sigma_1,\sigma_2,\cdots,\sigma_{L-1},\sigma_L\rangle$, where the values of $\sigma_i$ are
denoted $\pm$ for simplicity.
At zero magnetic field $H$ the two states with all spins positive $|++\cdots++\rangle$
or negative $|--\cdots--\rangle$ have the same energy. The extra magnetic field term
splits both states and the energy per spin is

\begin{equation}
\epsilon_{1,2} = -J (2- \frac{1}{L}) \pm H.
\label{en12}
\end{equation}

Assuming $H>0$ the $|++\cdots++\rangle$ state is always the singlet ground state. In order 
to find the first
excited states we have to flip the first or the last column ($i=1,L$) in the previous configurations.
In this way we get the four states $|-+\cdots++\rangle$, $|++\cdots+-\rangle$, $|+-\cdots--\rangle$
and $|--\cdots-+\rangle$. The magnetic field splits this level into two doublets and for the two 
first states their energy decreases when the $H$ increases according to the equation

\begin{equation}
\epsilon_{3,4} = -J (2- \frac{3}{L}) - H (1-\frac{2}{L}).
\label{en34}
\end{equation}

Therefore, we expect the crossing of the singlet state $|--\cdots--\rangle$ with
the doublet $|-+\cdots++\rangle$, $|++\cdots+-\rangle$ at a certain value of the bulk magnetic
field

\begin{equation}
H_x(T=0;L)= \frac{J}{L-1}.
\label{hxt0}
\end{equation}

Note, that for $T \to 0$ Eq.(\ref{1ordline}) reduces to Eq.(\ref{hxt0}) provided 
$A(T) \to J$ and $B(T),C(T) \to 0$.

\begin{figure}[b]
\centerline{
\rotate[r]{\psfig{file=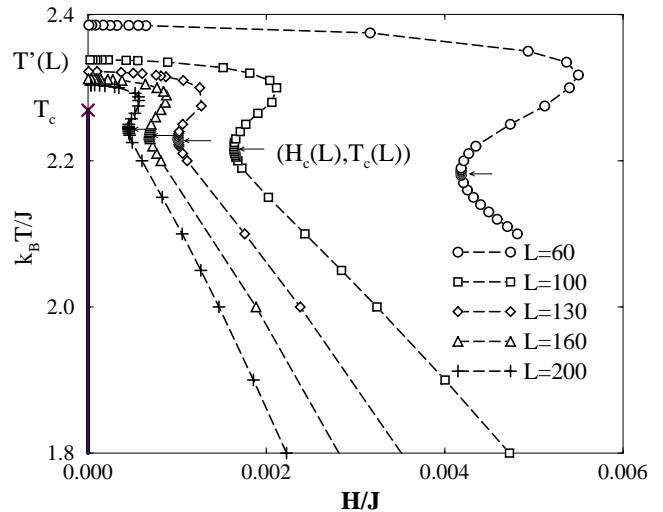,height=8.5cm}}}
\vskip 0.2truecm
\caption{The lines of coexistence of the excited states for different strip width $L$.
$T_c$ is the bulk critical point, whereas the thick
solid line denotes the bulk first-order line.
The arrows point at the inflection points where $H_x(T;L)$ has
a local minimum as a function of the temperature. $T'(L)$ describes
an end point where $H_x(T;L)$ ends on the $H=0$ axis.
The dashed lines are as guides for eye.}
\label{FIG1}
\end{figure}

At finite temperatures we do not have any more real crossing points but, so-called 
``the regions of avoided level crossing''. At $H=0$ the first two levels are separated as
$f_1-f_0 \sim \exp(-\sigma_0 L/k_B T)$,
so they are asymptotically degenerated for $L \to \infty$. The region of avoided level
crossing continues for nonzero magnetic fields up to $|H| \sim \exp(-\sigma_0 L/k_B T)$.
When we are interested in the behavior of $\xi$, we have to consider the second and third 
eigenvalues of the transfer matrix \cite{PF,APU}, where asymptotic degeneracy is also 
present for $f_1$ and $f_2$ \cite{NS}.
It is assumed that the difference $f_2-f_1$ and the avoided level crossing region centered
on $H_x$ are of order $\exp(-CL)$, where the coefficient C may be $H$ and $T$ dependent.
It makes sense to discuss algebraic shift of the value $H_x(T;L)$
for $L \to \infty$. In order to find out the value of $H_x$ in finite
temperatures at fixed $L$, we identify $H_x$ with the value of the bulk magnetic 
field where the second free-energy level $f_1$ has a maximum and the separation $f_2-f_1$
is the smalest one. The curves $H_x(T;L)$ 
present the coexistence of the excited states and are shown in Fig.2.

The curves indicate the phase boundaries between the two phases with different 
dependencies of $\xi$ on $L$ and $H$. As $L \to \infty$ the coexistence
lines of the excited states shift towards the $H=0$ axis that is intuitively 
clear at $T=0$. Since the width of the strip increases the energy of configurations 
where only one column of spins is flipped (Eq.(\ref{en34})) are close and close to 
the energy of configurations with all spins pointed in one direction (Eq.(\ref{en12})), so
in the $L \to \infty$ limit one has $H_x=0$. 

Let us discuss the scaling of the coexistence line $H_x(T;L)$ to the bulk first-order 
line (Fig.(\ref{FIG1})). To verify the bubble model predictions (Eq.(\ref{1ordline})) we have 
calculated series of values of $H_x(T={\rm const};L)$ for $L=20,40,\ldots,200$ and for 
temperatures ranging from $T\approx 0.44T_c$ up to $T\approx 0.99T_c$.

\vskip.4cm
\centerline{{\bf Table.} Scaling exponents of $H_x(T;L)$}
\centerline{extrapolated from the DMRG data.}
\centerline{
\begin{tabular}{|c|c|c|c|}
\hline\hline
$T$ & $\alpha$&  $\gamma$+1&  $\delta$+5/3  \\
\hline
1.00 & -0.9994(5) & -0.668(1) & -0.66(4) \\
1.50 & -0.9990(5)& -0.667(1) & -0.64(1) \\
1.75 & -1.000(1)& -0.668(2) & -0.64(1) \\
2.00 & -0.998(1) & -0.667(1)  & -0.67(1) \\
2.15 & -1.028(3) & -0.67(1) & -0.67(2) \\
2.20 & -1.002(6) & -0.69(1)  & -0.68(3) \\
\hline\hline
\end{tabular}
}
\vskip.6cm

Table shows the values of scaling exponents obtained from the DMRG data. Using the powerful 
extrapolation technique, the Bulirsch and Stoer (BST) method \cite{HenSch}, we obtained
an excellent agreement with Abraham {\em et al.}.

In order to get the $A$ coefficient in Eq.(\ref{1ordline}) one can compare Eqs. (\ref{P}) and 
(\ref{RS}). They have to agree at the value $H=H_x$ in the termodynamic limit, which implies
the following relation \cite{PS}:

\begin{equation}
A(T)=\sigma_0(T)/m(T).
\label{A}
\end{equation}

\begin{figure}[b]
\centerline{
\rotate[r]{\psfig{file=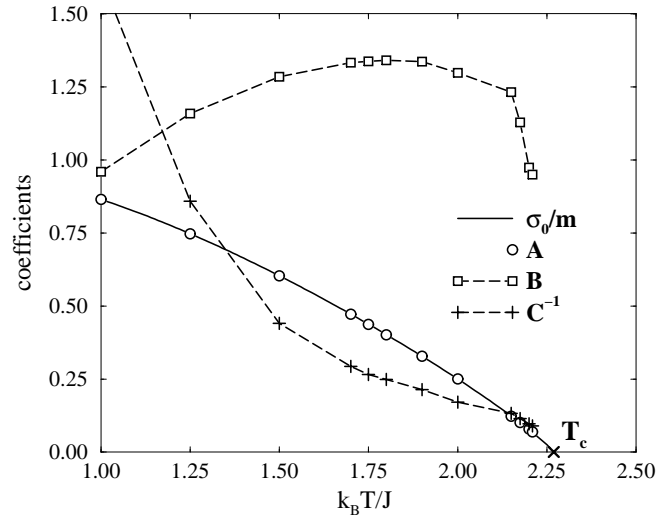,height=8.5cm}}}
\vskip 0.2truecm
\caption{The coefficients of the scaling of the coexistence line $H_x(L;T)$ to the bulk
first-order line. Solid line denotes the analytical result determined in the bubble model.
The symbols describe our numerical results. The dashed lines are as guides for eye.}
\label{FIG2}
\end{figure}

In Fig.\ref{FIG2} our data reconstruct this 
curve very well. To the best of our knowledge, the coefficients $B(T)$ 
and $C(T)$ in Eq.(\ref{1ordline}) have not been yet determinated, but our numerical 
results can predict their temperature behavior.

Close to $T_c$ the validity of Eq.(\ref{1ordline}) is limited because the scaling of points
of the $H_x$ curve is governed by the bulk critical point.
In order to study it in detail we have considered characteristic points of the upper
part of the $H_x$ curve: the inflection points ($H_c(L)$,$T_c(L)$) and the end 
points $T'(L)$ (see, Fig.(\ref{FIG1})), where the following scaling is expected:

\begin{eqnarray}
&& \tau_c(L)=(T_{\rm c} - T_{\rm c}(L))/T_{\rm c} \sim L^{-y_T},
\nonumber \\
&& H_{\rm c}(L) \sim L^{-y_H}.
\label{shiftTH}
\end{eqnarray}
Here, $y_T=1$ and $y_H=15/8$ are the thermal and magnetic exponents of the two-dimensional 
Ising model.

To verify the scaling to the critical point ($H=0$,$\tau=0$) 
we found out the inflection points and the end points for 
$L=30,60,100,130, 160$ and $200$ using subsequently the BST technique. We have examined 
the scaling form (\ref{shiftTH}) for $L \to \infty$ and found very good agreement

\begin{eqnarray}
&& \tau_c =0.00006(6), \,\,\,\,\quad {\rm and} \,\, y_T =1.005(5)
\nonumber \\
&& H_c =-0.0006(6), \quad {\rm and} \,\, y_H =1.876(8)
\nonumber \\
&& \tau' =0.0000(3), \,\,\,\,\,\,\,\quad {\rm and} \,\, y_T =1.006(6)
\nonumber
\end{eqnarray}

Note, that for $T \to T_c$ and $L \to \infty$ we can reproduce the scaling form (\ref{shiftTH}) 
from Eq.(\ref{1ordline})
by assuming that $A(T) \to 0$, $B(T) \to 0$, and $C(T) \to \infty$. This is in agreement with
our numerical estimations for scaling coefficients as depicted in Fig.\ref{FIG2}. Of course, this
relation is not valid at $T_c$.

\begin{figure}[b]
\centerline{
\rotate[r]{\psfig{file=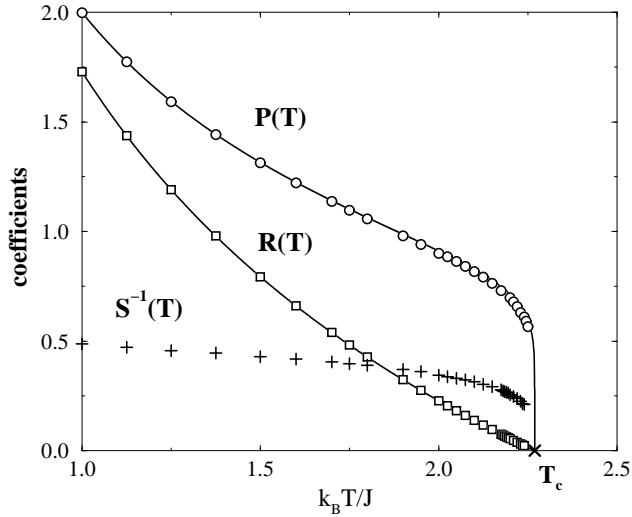,height=8.5cm}}}
\vskip 0.2truecm
\caption{The coefficients of the correlation length in Eqs.(\ref{P},\ref{RS}).
Solid lines denote the bubble model results: $P(T)=2m/k_{B}T$ and $R(T)=2\sigma_{0}/k_{B}T$. 
The symbols describe the corresponding DMRG results.}
\label{FIG3}
\end{figure}

In order to analyze the behavior of the correlation length we have derived $1/\xi$  for $L$ between
$100$ and $300$ in temperatures below $T_c(L)$. To examine the form of Eq.(\ref{P}) 
firstly we have confirmed the linear dependence of the coefficient on $L$. Next we compared 
our numerical results with the coefficients $P(T)$ and $R(T)$ in Eqs.(\ref{P},\ref{RS}). What 
is more, we presented the temperature dependence of the $S(T)$ coefficient which was not 
determined in the bubble model (see, Fig.\ref{FIG3}).

When temperature raises, more and more complex configurations on the Ising strip 
contribute to the free energy in contrast to the assumption of the bubble model \cite{APU}. 
Consequently, in high temperatures the validity of Eqs. (\ref{P},\ref{RS}) is limited to 
a narrow range of $H$. The bubble model predictions are also spoiled by the presence of strong 
bulk magnetic field. That is why, the higher is temperature the smaller $H$ is necessary 
to recover the linear dependence of $1/\xi$ on $H$, as in Eq.(\ref{P}). Similarly, when 
$T \to T_c(L)$ the regime with the $H^{2/3}$ dependence of $1/\xi$ (Eq.(\ref{RS})), close 
to the right side of the coexistence line, shrinks to zero.

In conclusion, we have used the density-matrix renomalization-group method
to obtain reliable information about the two-dimensional Ising model in
the bulk magnetic field.
We have confirmed the crossover related to the correlation length analyzed
before for the bubble model \cite{APU}.
Our study has not been limited to subcritical temperatures and small bulk fields. 
We have confirmed Abraham {\em et al.} predictions for the scaling of the
first-order line in the subcritical region. Morover, we have established
the precise scaling form for the bulk magnetic field by numerically determining 
coefficients $B(T)$ and $C(T)$ in Eq.(\ref{1ordline}).
Furthermore, we have extended the analysis of the bubble model to {\em critical
region} verifying that the scaling behavior is governed by the bulk critical point.
Finally, we numerically confirmed the magnetic field dependence of the correlation length,
simultaneosly extracting previously unknown coefficient $S(T)$ in Eq.(\ref{RS}).
Above results demonstrate that for two-dimensional classical systems the DMRG technique 
provides significantly accurate 
data for studying equilibrium properties of large systems in a {\em nonvanishing} bulk magnetic
field.

\acknowledgments

I thank A.O. Parry for suggesting me the topic of this 
Raport. I am grateful to E. Carlon and T.K. Kope\'{c} for critical reading of the manuscript.
This work was supported by the Polish Science Committee (KBN) under Grant No. 2P03B10616.

\end{multicols}
\end{document}